Hydrophilic Porous Materials as Helmet Padding Able to Prevent Traumatic Brain Injuries

J. B. Sokoloff, Physics Department, Northeastern University, Boston, MA 02115,

E-mail address: j.sokoloff@neu.edu

**Abstract**

The ideal way to reduce the likelihood of traumatic brain injuries for a player engaged in an impact sport such as football or hockey, as a result of an impact of his/her helmet with a surface or another player would be to reduce the average value of the acceleration of the player's head in an impact as much as possible. The minimum possible value of the average deceleration of the head is inversely proportional to the helmet padding thickness. Since there are practical limits to its maximum thickness, it is difficult to significantly reduce the average acceleration. There is evidence, however, that the peak, rather than the average, acceleration is the most significant cause of brain injury. It is proposed here that brain injuries that occur as a result of an impact, could be reduced by using as padding a hydrophilic porous material swollen with fluid. The friction experienced by the fluid as it is squeezed out of the porous material in an impact can "tune" the acceleration of the skull so that it is never significantly higher than its average value during the impact.

I. Introduction

In Ref. 1, Rowson and Duma conducted a study dedicated to determining the probability of the occurrence of concussions among football players during a game. They put forward a function of the peak deceleration $a_{max}$ which gives the probability of the occurrence of a concussion for a given peak deceleration during an impact, based on a logical regression analysis of injury rates on the football field correlated with drop tests done on helmets. Their function is

$$R(a'_{max}) = \frac{1}{1+e^{-(\alpha+\beta a'_{max})}}, \qquad (1)$$

where $a'_{max} = a_{max}/g$, i.e., $a_{max}'$ is the deceleration $a_{max}$ in units of the gravitational acceleration $g$. Using statistics from NCAA college games, they obtain the following values for the parameters: $\alpha = -9.805$ and $\beta = 0.051$. Using statistics for NFL games these parameters have the values: $\alpha = -9.828$ and $\beta = 0.0497$. Although the study reported in Ref. 1 only dealt with linear acceleration, it can be argued that the same forces on parts of the brain resulting from large linear acceleration of the brain, that are responsible for the occurrence of a concussion, will also occur for rotational acceleration that occurs in a rotational impact. This argument is based on the fact that the maximum value of the peak linear acceleration associated with a peak angular acceleration $\alpha_{max}$ is given by $a_{max} = r_0 \alpha_{max}$, where $r_0$ is the distance between the center of mass of the brain and a point just below the inside surface of the skull[2]. This argument tells us that the way to reduce the probability of occurrence of a concussion in a rotational impact is to reduce the peak value of the angular acceleration.

The ideal way to protect the brain from injury during an impact is to reduce the deceleration of the head as much as possible[3-16]. Assuming as a first approximation that during an impact of a helmeted head with a hard surface, the helmet stops moving instantaneously and the padding compresses by an amount $\Delta z$ as it reduces the speed of the head to zero, the average deceleration can be found from

$$V_0^2 + 2a_{av}\Delta z = V^2 = 0, \qquad (2)$$

where $V_0$ and $V$ are the initial and final velocities, respectively, and $a_{av}$ is the average deceleration (average over $\Delta z$). Then, it follows that

$$a_{av} = -\frac{V_0^2}{2\Delta z}. \qquad (3)$$

For example, if the velocity just before the impact $V_0 = 5m/s$ (a typical running speed which corresponds to an impact energy of $37.5J$, assuming the mass of the head to be $3kg$) and $\Delta z = 2.5 \times 10^{-2} m$, $a_{av} = 500 m/s^2$. Then, it appears that ideally the padding should be made as soft as possible, while still stiff enough or thick enough so that it does not get completely compressed in an impact. There are practical limits on how thick one can make the padding (to prevent the padding from completely compressing before the velocity of the head is reduced to zero), however, no matter what material is used. What is proposed here is that since there are practical limitations on how small one can make the average deceleration because of limitations on the padding thickness, an alternative approach is to design the padding so that the peak deceleration is never significantly larger than its average value.

The padding used in football helmets today is either a polyurethane or vinyl nitrate foam, which has the disadvantage that it results in a very large peak deceleration[17]. As was stated earlier, for a given pad thickness, the lowest average deceleration that can be achieved is $V_0^2/(2\ell)$, where $\ell$ is the thickness of the padding. For foam used in typical football helmets for an impact energy of $30J$, which corresponds to an initial velocity of $4.47m/s$, the maximum acceleration for the foam padding in Ref. 17 was 408g. This gives a value for $R$, given in Eq. (1), almost equal to one, implying a high probability of a brain injury. Even the VICIS helmet, which is considered the state of the art in injury preventing football helmets, has a highly peaked deceleration profile. Fig. 1 shows the deceleration versus time for the VICIS helmet found in finite element simulations performed by the NFL to reproduce results of impact tests done on several helmets presently in use[18]. Similar time dependence for the deceleration was found for the other helmets that they tested. The value of $R$ for the data shown in Fig. 1 is 0.0231, using the parameters given above for NFL games. The authors of Ref. 17 proposed using shear thickening fluids (STF) as helmet padding. These are colloid materials whose viscosity increases rapidly as the rate at which stress is applied to them increases. For a $3.3cm$ thick pad of the STF studied in Ref. 17, the maximum acceleration for a $30J$ impact was $119g$, which gives $R = 0.02$. We will see that padding made from a 4cm thick hydrophilic porous material swollen with water can give a peak deceleration of $667m/s^2 = 68.1g$ with the optimum choice of parameters, for an initial velocity of $5m/s$, which gives *R=0.00159.* One reason that porous materials swollen with water can potentially do a better job in preventing brain injuries is that such materials provide significant deceleration at the beginning of the impact (i.e., well before the padding is significantly compressed). This makes it easier to produce a deceleration during the impact close to being constant. Ref. 19 suggests that nearly constant acceleration could be produced by expulsion of water from a water shock absorber (i.e., a container with a hole in it that allows water to escape as the container is compressed during an impact to it). Porous materials swollen with water can produce close to the constant acceleration that such a shock absorber can produce, but as will be discussed in section III, the porous material may be better suited to be used as helmet padding[20] because it might be easier to put it in a helmet.

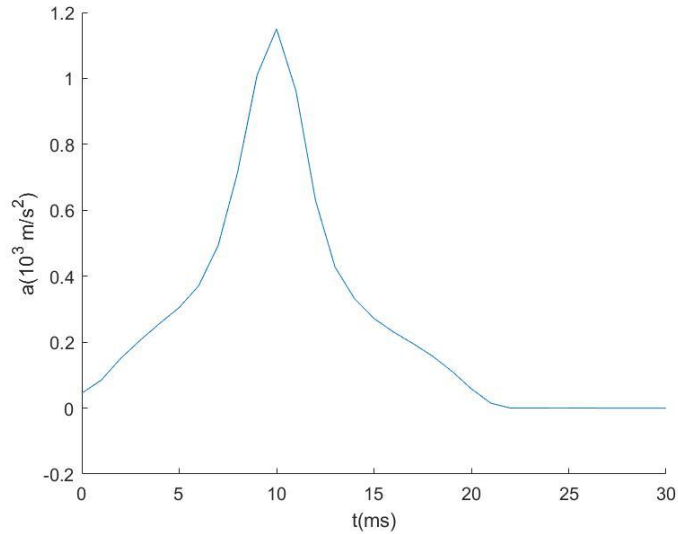

Figure 1: The acceleration of a model head in a VICIS helmet as a function of time from the NFL's simulation based on their testing of this helmet, for an initial velocity of the helmet of *3.61m/s* and a final velocity of  *- 3.61m/s.*

### II.     Theory of Porous Material Helmet Padding

Consider a porous padding material, which is initially swollen with water, when it is impacted with a slightly curved object of mass $m$ and a velocity $V_0$. This is a model for a head inside a helmet making an impact with a pad whose dimensions are small compared of those of the head, eliminating the need to account for the head's curvature, to a good approximation. Let us enclose the porous material in a circular cylinder with holes in its side, to allow fluid to escape as the material is compressed, but with a solid bottom and piston, so that the fluid flows only out the sides, as illustrated in Fig. 2. The cylinder must be enclosed in a watertight soft elastic wrapper to catch fluid that is expelled during compression, keeping it in contact with the sides of the cylinder and thus allowing it to be reabsorbed after the impact.

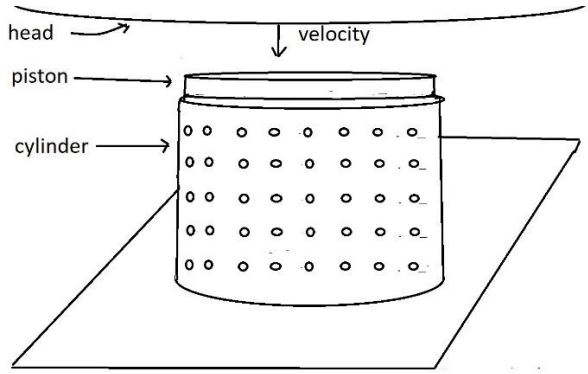

Figure 2: This figure illustrates how a porous material can be used to function as helmet padding. The porous material is contained within the cylinder. There would be several of these pads distributed around the inside of the helmet.

A major source of head injury is believed to be due to rotational acceleration of the head. In the case of head rotation, it is the maximum angular acceleration that must be reduced to prevent concussions, as described above. The maximum angular acceleration can be reduced by including in the helmet impact absorbing cylinders described above, but attached to the sides of an insert in the helmet (in contact with the head), in order to reduce the maximum angular acceleration of the head in a rotational impact, as illustrated in Fig.3.

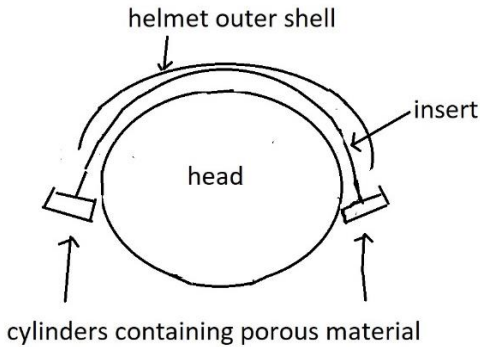

Figure 3: A schematic of an insert placed in the helmet to mitigate effects of rotational impacts. The insert shell would actually be in contact with the head.

Let the initial and instantaneous thickness of the porous material be $\ell$ and $z$, respectively, and hence, $\Delta z$ in Eqs. (2) and (3) is equal to $\ell - z$. The force that the surface of the porous material must exert on the head to decelerate it is given by Newton's second law,

$$F = m\frac{d^2 z}{dt^2}, \qquad (4)$$

where *m* is the mass of the head. Then setting the rate of decrease of the volume of the material equal to the volume rate of flow of fluid through the sides, we get

$$\pi R^2 \frac{dz}{dt} = 2\pi R z v, \qquad (5)$$

where $v$ is the mean velocity of fluid flow out the sides of the cylinder, and as discussed above, we have assumed that the pad is enclosed in a circular cylinder of radius $R$. Here, it will be assumed that $F$ is dominated by the friction that the water exerts on the porous material when it is expelled from it. This assumption will be justified later by numerically solving a model that includes both this friction mechanism and the elasticity of the solid material in the padding. Therefore, we may assume that the flow velocity of the water $v$ is determined by the permeability of the porous material, and hence, is given by

$$v \sim k(F/\pi R^2)(1/R) \quad (6)$$

where $k$ is the permeability of the material, $F/\pi R^2$ is the fluid pressure applied to the porous material and the mean distance that the expelled fluid flows is of the order of $R$. (The permeability is often expressed as a quantity with units of distance squared, given by $\eta k$ in terms of the permeability $k$ defined by Eq. (6), where $\eta$ is the dynamic viscosity of water.)

For the geometry shown in Fig. 2, the permeability should be proportional to the sample thickness, and hence, $k = k_0(z/\ell)$. To justify this assumption, think of the water flowing through a bunch of pipes running from inside the porous material to the outside. Then since only the vertical dimension of each pipe gets reduced by a factor $z/\ell$, the total cross-sectional area of each pipe gets reduced by this factor. Combining Eqs. (4-6), we obtain the differential equation:

$$\frac{d^2 z}{dt^2} = -\frac{\ell \pi R^4}{2k_0 z^2 m}\frac{dz}{dt} = -\frac{a_0 \ell^2}{V_0 z^2}\frac{dz}{dt}, \quad (7)$$

where $a_0$ is the initial acceleration. The elasticity of the solid part of the porous material can be included by adding a term $-(G\pi R^2/m)(1-z/\ell)$ to the right hand side of Eq. (7), where $G$ is the compressional elastic constant of the solid part of the porous material. For now, we will neglect this term. As mentioned above, however, it will be demonstrated below that the results obtained without including this term are not modified significantly if it is included. There exist hydrogels with cavities which can have a different dependence of $k$ on $z/\ell$ [21]. Results of calculations done on this model with $k = k_0(z/\ell)^\alpha$ for a few values of $\alpha$ are given in the appendix. Integrating Eq. (7) once, assuming that the initial value of $dz/dt = -V_0$, we obtain

$$\frac{dz}{dt} = -V_0 + (a_0 \ell / V_0)(\frac{\ell}{z}-1), \quad (8)$$

whose solution is

$$\ln\left|\frac{A\ell}{(1+A)z-A\ell}\right| = \frac{V_0(1+A)^2 t}{\ell A} + (\ell - z)\frac{1+A}{\ell A}, \quad (9)$$

where $A = a_0 \ell / V_0^2$. As $t$ becomes large compared to

$$\frac{\ell}{V_0}\frac{A}{(1+A)^2}, \qquad (10a)$$

the right hand side of Eq. (9) becomes large, and therefore, the value of z decays from $\ell$ to its final value $z_{min}$, given by

$$\frac{z_{min}}{\ell} = \frac{A}{1+A} \qquad (10b)$$

(since the logarithm becomes infinite at this value). Substituting Eq. (8) into Eq. (7), we obtain

$$a = \frac{d^2 z}{dt^2} = a_0 \frac{1}{u^2}\left[1 - A\left(u^{-1} - 1\right)\right], \qquad (11)$$

where $u = z/\ell$. The value of $z$ at which $a$ is maximum ($z_{a,max}$) is obtained from

$$\frac{1}{\ell^2 a_0}\frac{da}{dz} = -\frac{2}{z^3}\left[1 - A\left(u^{-1} - 1\right)\right] + A\frac{1}{uz^3} = 0, \qquad (12)$$

whose solution is

$$\frac{z_{a,max}}{\ell} = \frac{3}{2}\frac{A}{1+A} = \frac{3}{2}\frac{z_{min}}{\ell}. \qquad (13)$$

Substituting $z_{a,max}$ for $z$ in Eq. (11), we obtain for the maximum acceleration

$$\frac{a_{max}}{a_0} = \frac{4}{27}\frac{(1+A)^3}{A^2}. \qquad (14)$$

Since

$$z_{a,max} = (3/2)z_{min} < \ell, \qquad (15)$$

(because $\ell$ is the maximum value of z), we find from Eq. (13) that we must have

$$A \leq 2. \qquad (16)$$

If this condition is not satisfied, $a_{max} = a_0$, which occurs right at the beginning of the impact, when $z = \ell$. The average (over $z$) of the acceleration is given by

$$a_{av} = \frac{V_0^2}{2(\ell - z_{min})} = a_0\left(\frac{1+A}{2A}\right), \qquad (17)$$

from Eq. (3) (noting that $\Delta z$ in Eq. (3) is equal to $\ell - z_{min}$) and Eq. (10b), and hence, from Eqs. (14) and (17)

$$\frac{a_{max}}{a_{av}} = \frac{8}{27}\frac{(1+A)^2}{A}, \qquad (18)$$

for $A \leq 2$. If we minimize Eq. (18) with respect to $A$, we find that the minimum value of $a_{max}/a_{av} = 1.19$, which occurs for $A = 1$.

Let us now look at how all of the above parameters depend on $\ell$. Since

$$A = \frac{a_0 \ell}{V_0^2} = \frac{\pi R^4}{2k_0 m V_0}, \qquad (19)$$

we see that $A$ is independent of $\ell$, and hence, $z_{min}/\ell$ and $z_{a,max}/\ell$ are independent of $\ell$. From the definition of $A$, given above Eq. (10a), along with Eqs. (14) and (17), we find that the quantities $a_0$, $a_{max}$ and $a_{av}$ are inversely proportional to $\ell$. The table below illustrates how the ratio $a_{max}/a_{av}$ depends on $A$

**Table I**

| A | $a_{max}/a_{av}$ |
|---|---|
| 0 | ∞ |
| 0.5 | 1.33 |
| 1.0 | 1.19 |
| 1.5 | 1.23 |
| 2.0 | 1.33 |
| 3.0 | 1.5 |
| 4.0 | 1.6 |
| ∞ | 2.0 |

Table I: The ratio $a_{max}/a_{av}$ is given for several values of A.

In this table, $a_{max}/a_{av} = a_0/a_{av}$ for $A > 2$. It illustrates that for $0.5 < A < 2$, $a_{max}/a_{av} \leq 1.33$. Since from Eq. (19), $V_0 = (\pi R^4 / 2k_0 m A)$, we conclude that such low values of $a_{max}/a_{av}$ can be obtained for an impact velocity range of $0.5(\pi R^4 / 2k_0 m) < V_0 < 2(\pi R^4 / 2k_0 m)$, which would mean for example that if we chose the parameters so that $(\pi R^4 / 2k_0 m) = 5 m/s$, we would have $a_{max}/a_{av} \leq 1.33$ for $2.5 m/s < V_0 < 10 m/s$. The relative amount of compressions for a given value of A is given by $(\ell - z)/\ell = (1+A)^{-1}$, which for the above range of values of A, varies between 0.333 and 0.667. The values of $a_{max}/a_{av}$ given in the appendix for other dependencies of the permeability on the compression of padding are comparable to these values. This implies that there are probably several different porous materials that can be used to reduce $a_{max}/a_{av}$.

The above expressions for $a_{max}$ and $a_{av}$ can also be written as

$$a_{max} = \frac{4}{27} \frac{(1+A)^3}{A} \frac{V_0^2}{\ell} \qquad (20)$$

and

$$a_{av} = (1+A) \frac{V_0^2}{2\ell}. \qquad (21)$$

It is easily shown that $a_{max}$ is minimum for $A = 0.5$, and for this value $a_{max} = V_0^2/\ell$, $a_{av} = 0.75 V_0^2/\ell$ and $\ell - z_{min} = 0.667\ell$. Then, for example, if $V_0 = 5 m/s$ and $\ell = 3.75 \times 10^{-2} m$, $a_{av} = 500 m/s^2$ and $a_{max} = 667 m/s^2$.

The quantity $(\pi R^4 / 2k_0 m) = 5 m/s$ for $R = 3.65 cm$ for example, $m=3kg$ and $k_0 = 1.86 \times 10^{-7} m^4/Ns$. Since $k_0 = \zeta^2/\eta$, where $\zeta$ is the root mean square pore size, $\zeta = (k_0 \eta)^{1/2} = 1.35 \times 10^{-5} m$, where $\eta$ is the viscosity of water.

In order to justify the neglect of the elasticity of the solid material in the above discussion, let us include the elastic term discussed under Eq. (7), to account for the elasticity of the solid material of the sample and solve the resulting differential equation,

$$\frac{d^2 z}{dt^2} = -\frac{a_0 \ell^2}{V_0 z^2} \frac{dz}{dt} - \frac{G \pi R^2}{m} \left(1 - \frac{z}{\ell}\right), \qquad (22)$$

numerically. When the weight is placed at rest on the padding the material,

$$G \pi R^2 \left(1 - \frac{z}{\ell}\right) = mg. \qquad (23)$$

The results are shown in Fig. 4a for the value of G which gives $1 - z/\ell = 0.626$ when the weight rests on the padding, determined from Eq. (23). In Fig. 4b the value of G is 20 times larger than its value in Fig. 4a, and in Fig. 4c, G=0 for comparison. From these figures, it can be seen that including the elasticity term does not significantly change the acceleration versus compression curves, which indicates that the above calculations based on a model that assumes that the water friction dominates over the elasticity is a good approximation.

9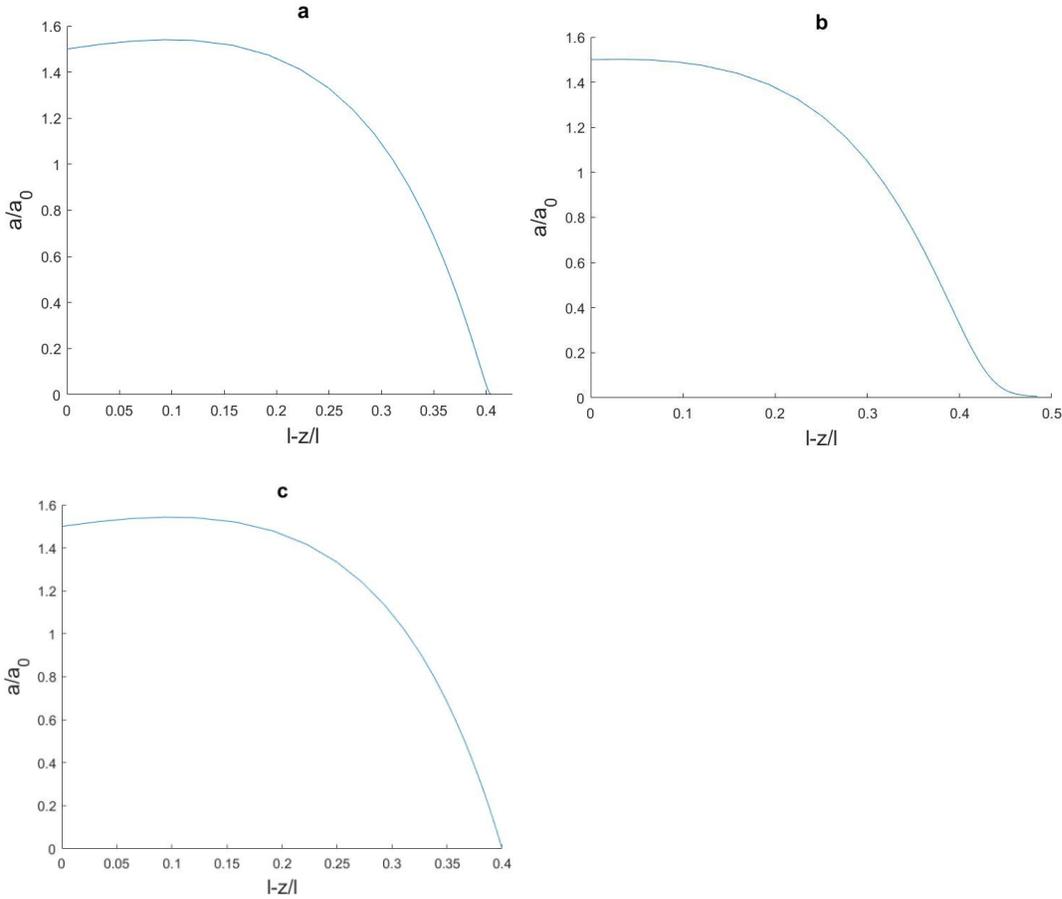

Figure 4: The acceleration of the weight is plotted as a function of the compression of the padding during the impact with *A=1.5* and for a. $G = 1.12 \times 10^4 Pa$ chosen so that when a *3kg* weight rests on the pad $(1 - z/\ell)$ given by Eq. (23) is equal to 0.626, b. *G* is 20 times the value in a, and c. for *G=0*.

Let us consider a numerical example of how effective the porous material is in preventing brain injuries. For the maximum acceleration of $667 m/s^2$ or 68.1g found in the example worked out under Eq. (20), we obtain from Eq. (1) that the probability of a concussion, $R = 0.00178$ for the NCAA parameters and for the NFL parameters, it is $R = 0.00159$, showing that the probability of a brain injury is extremely low.

### III.    A possible Way to Produce a Porous Material Helmet Pad which is Able to Prevent Brain Injuries

In order to achieve a maximum acceleration comparable to the relatively small values given above, we saw that the pore size required for our porous material must be of the order of between $10^{-6} - 10^{-5} m$. Recently, there has been an interest in developing highly porous materials using optical etching of ceramic materials [22-24] or 3d printing.[25-28]. The porous materials reported in Refs. [22-24] by Greer's group have compressional moduli of the order of $10^4 - 10^6 Pa$ and those produced by Yakacki's group from liquid crystal elastomers[25-28] give moduli of the order of $10^4 - 10^5 Pa$. The value of G which gives $z/\ell = 0.5$, for example, when the weight comes to rest after the impact, is in the range of compressional moduli that occur for the porous liquid crystal elastomers discussed above [25-28].

## IV. Comparisons with other Approaches

Studies of the head deceleration produced by a multilayer pad with a stiffer pad sandwiched between more compressible layers (padding used by the military) show that this type of padding produces a maximum deceleration rate of about 120g, which according to Eq. (1) gives a 10% probability of a concussion[29]. Padding consisting of a porous material swollen with water, as well as the shear thickening liquid padding, discussed in Ref. 17, as well as several other proposed methods that will be discussed in this section give a much lower probability of a concussion. Recently, it has been demonstrated that a shock absorber consisting of a collapsible container containing water can result in a low peak acceleration in an impact.[19]. This approach has the advantage that the shock absorber can be engineered to produce a precisely constant deceleration rate during an impact to it. It has the disadvantages, however, that it is difficult to design such a device that will fit into a helmet and it is not obvious that it is possible to design such a device that will re-absorb the water expelled in an impact before the next impact takes place. Also, as was noted in the introduction, it has been demonstrated that padding that contains a shear thickening liquid is also able to reduce the peak acceleration in an impact[17], although from the discussion in the previous sections, the mechanism proposed here can potentially be more effective at preventing concussions. More recently Mistry, et. al. [27] have shown that "soft elasticity" in liquid single domain crystal elastomers can give rise to a stress versus-strain curve which flattens out as the strain increases as a result of rotation of the liquid crystal directors during an impact. Such materials were shown to give rise to a stress versus strain curve which is a little closer to the ideal constant stress versus strain behavior[19], but the stress-strain curve that they found in their impact tests is still farther away from the behavior that is possible for the mechanism suggested in this paper. Spinelli, et. al.[30], have demonstrated that a helmet held on the head with stretchable straps containing a shear thickening liquid can produce a reasonably constant acceleration of the head in an impact, at least for impact speeds below $4.5 m/s$. At $4.5 m/s$, however, the ratio $a_{max}/a_{av}$ appears to be closer to 2.

## V. Conclusions

It was argued that helmet padding made from hydrophilic porous materials swollen with water with pore size of between 1 and 10 microns can be effective in preventing concussions in high impact sports by reducing the value of the maximum deceleration and angular deceleration rate of the head relative to the helmet during an impact to the helmet. Although there is evidence that even repeated sub-concussive impacts cause long term damage to the brain[31-37], reducing the peak acceleration, which reduces the maximum forces acting on various parts of the brain will, in addition to reducing the likelihood of a concussion, likely also reduce the damage caused by sub-concussive impacts.

**Appendix**

Results of several calculations of the acceleration versus the amount of compression $1 - z/\ell$ of a porous material swollen with water for a permeability of the form $k = k_0 (z/\ell)^\alpha$ with a few values of $\alpha$ will be given here. Eq. (7) is replaced by

$$\frac{d^2z}{dt^2} = -\frac{a_0}{V_0}\left(\frac{\ell}{z}\right)^{\alpha+1}\frac{dz}{dt} \qquad (A1)$$

whose solution for the velocity is

$$\frac{dz}{dt} = V_0\left[-1 + A(u^{-\alpha} - 1)\right] \qquad (A2)$$

where $u = z/\ell$. Substituting Eq. (A2) into Eq. (A1), we get

$$a = \frac{d^2z}{dt^2} = a_0\left[\left(1 + \frac{A}{\alpha}\right)\frac{1}{u^{\alpha+1}} - \frac{A}{n}\frac{1}{u^{2\alpha+1}}\right]. \qquad (A3)$$

The average acceleration is given by

$$a_{av} = \frac{V_0^2}{2(\ell - u_{min})} = \frac{a_0}{A}\frac{1}{1 - u_{min}}, \qquad (A4)$$

where $u_{min}$ is the value of $u$ at which $dz/dt=0$. Based on Eq. (A3), $a$ is plotted as a function of $u$ for $\alpha = 1.5$ for some values of $A$ below:

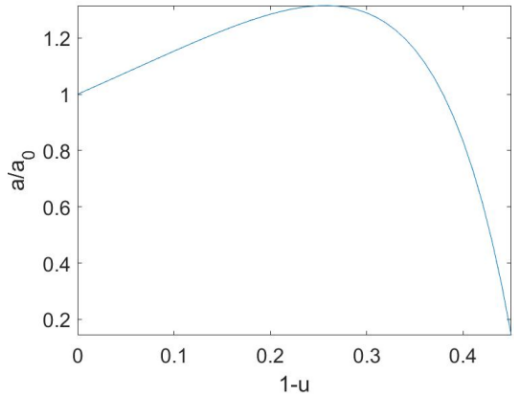

Figure 5: This is a plot of $a/a_0$ versus 1-u for $\alpha = 1.5$ and for A=1.

From Eq. (A4), $a_{av}/a_0 = 1.00$, so that $a_{max}/a_{av} = 1.3$.

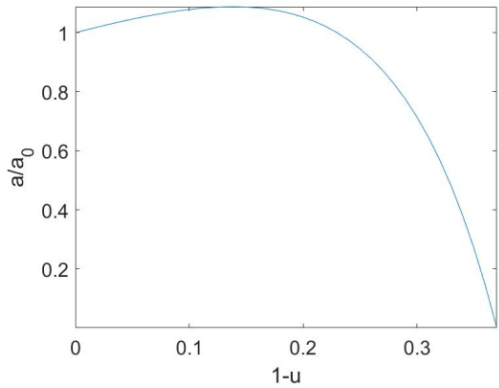

Figure 6: A plot of *a* versus *1-u* for $\alpha = 1.5$ and for *A=1.5*.

From Eq. (A4), $a_{av}/a_0 = 0.952$, so that $a_{max}/a_{av} = 1.16$.

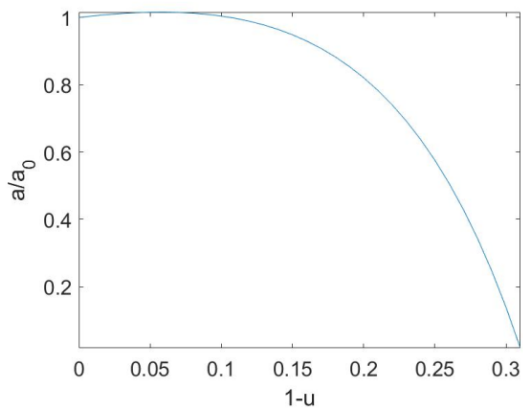

Figure 7: A plot of *a* versus *1-u* for $\alpha = 1.5$ and *A=2*.

From Eq. (A4), $a_{av}/a_0 = 0.833$, so that $a_{max}/a_{av} = 1.20$.

Plots are given below of a versus 1-u for $\alpha = 3$ for several values of A:

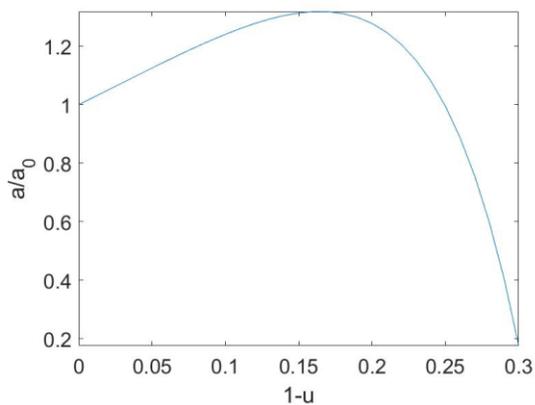

Figure 8: A plot of *a* versus *1-u* for $\alpha = 3$ for *A=1_5*.

From Eq. (A4), $a_{av}/a_0 = 1.11$, so that $a_{max}/a_{av} = 1.17$.

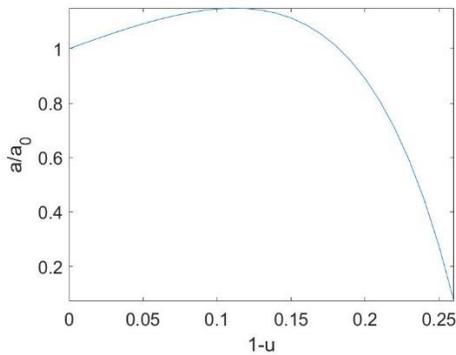

Figure 9: A plot of *a* versus *1-u* for $\alpha = 3$ and *A=2*.

From Eq. (A4), $a_{av}/a_0 = 1.00$, so that $a_{max}/a_{av} = 1.1$.

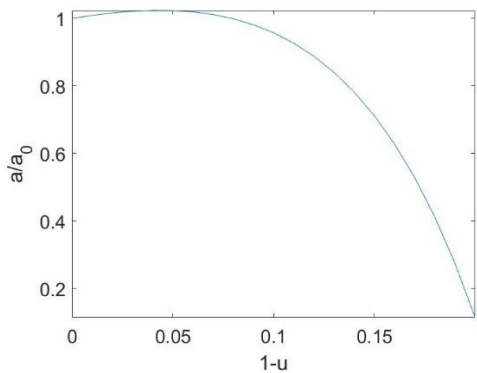

Figure 10: A plot of *a* versus *1-u* for $\alpha = 3$ and *A=3*.

From Eq. (A4), $a_{av}/a_0 = 0.809$, so that $a_{max}/a_{av} = 1.24$.

### Ackowledgements

I wish to thank Robert Weiss and Brian Vogt of the University of Akron and Adam Ekenseair of Northeastern University for discussions that we had related to this project. Regarding the material quoted in Ref. 18: The author(s) acknowledge Biomechanics Consulting and Research, LC and Football Research, Inc. for supporting the development and licensing the use of models used in this study. The study's conclusions do not represent the conclusions or opinions of either such entity.